\documentclass[twocolumn]{aastex62}
\usepackage{lineno}
\usepackage{amsmath}
\usepackage{lipsum}
\usepackage{blindtext}
\usepackage{graphicx} 
\usepackage{changepage}
\usepackage{array}
\usepackage{multirow}
\usepackage{tabularx}
\let\bfseries\mdseries

\begin{document}
\title{\textbf{A Catalog of Broad H$\alpha$ and H$\beta$ Active Galactic Nuclei in MaNGA}}
\shorttitle{Broad Line AGN Detection}
	\shortauthors{Negus et al.}

\author[0000-0003-2667-7645]{James Negus}
\email{$^{\star}$ james.negus@colorado.edu}
\affil{University of Colorado Boulder, 2000 Colorado Avenue, Boulder, CO 80309, USA}

\author[0000-0001-8627-4907]{Julia M. Comerford}
\affil{University of Colorado Boulder, 2000 Colorado Avenue, Boulder, CO 80309, USA}

\author[0000-0002-2713-0628]{Francisco M\"uller S\'anchez}
\affiliation{University of Memphis, 3720 Alumni Avenue, Memphis, TN 38152, USA}

\begin{abstract}
Broad H$\alpha$ and H$\beta$ emission lines (FWHM $>$ 1,000 km s$^{-1}$) are incredibly efficient tracers of the high-velocity clouds encircling Active Galactic Nuclei (AGN). As a result, we search for these broad line AGN in the Sloan Digital Sky Survey's Mapping Nearby Galaxies at Apache Point Observatory (MaNGA) catalog. We identify 301 broad-line H$\alpha$ galaxies and 801 broad-line H$\beta$ galaxies in the catalog. In total, we detect 1,042 unique broad-line galaxies with luminosities between 10$^{37}$ - 10$^{43}$ erg s$^{-1}$; 60 feature both broad H$\alpha$ and broad H$\beta$ emission. We also determine that the broad line region radius ranges between 0.01 - 46 light days, with a median radius of 0.1 light days (0.02 pc) for our broad H$\beta$ sample. In addition, we find that both samples feature a higher fraction of galaxy mergers (44\% for the broad H$\alpha$ sample and 43\% for the broad H$\beta$ sample), compared to the full MaNGA galaxy sample (26\%), which suggests that merger-driven fueling is strongly active in our sample. 

\end{abstract}


\section{Introduction} \label{sec:intro3}

The search for large populations of Active Galactic Nuclei (AGNs) in surveys has been extensive in recent years, and these efforts have utilized multiple bands of the electromagnetic spectrum (e.g., Radio, IR, Visible, and X-ray; e.g., \cite{2017FrASS...4...35P, 2020ApJ...901..159C}). A primary motivation for assembling catalogs with robust populations of AGNs is to better decipher the relationship between central supermassive black holes (SMBHs; $M_{\rm{BH}} > 10^{6} M_{\odot})$ and their parent galaxies (e.g., the M-$\sigma$ relationship; \cite{2000ApJ...539L..13G, 2000ApJ...539L...9F, 2005MNRAS.364.1337S, 2008ApJS..175..390H}). SMBH mass has also been shown to correlate tightly with the mass and luminosity of the host galaxy's galactic bulge (e.g., \cite{2003ApJ...589L..21M, 2011MNRAS.412.2211G, 2012MNRAS.419.2497B, 2020ApJ...888...37D}).

Additionally, several authors present evidence that AGN feedback effectively suppresses star formation (e.g., \cite{1998A&A...331L...1S, 2004MNRAS.355..995D, 2005MNRAS.361..776S, 2006MNRAS.365...11C, 2006ApJ...652..864H, 2006ApJ...653...86T, 2008MNRAS.388..587L, 2008MNRAS.391..481S, 2008MNRAS.385.1846M, 2011ApJ...739...69M, 2012ARA&A..50..455F}), while others contend that it may stimulate star formation in certain regimes (e.g., \cite{2005MNRAS.364.1337S, 2010ApJ...709.1018V}). However, the true nature of these processes remains elusive. To address this, extensive samples of AGNs can serve to shed light on how well-defined AGN parameters, like AGN luminosity and SMBH mass, correlate to features of the host galaxy, which include but are not limited to: morphology, the presence of a companion, and/or stellar properties. 

Broad emission lines (BLs) in the optical and UV bands can help ensure accurate AGN identification, as they can serve as the fundamental signatures of an AGN. For AGNs, these lines are produced in a high-density region, termed the broad line region (BLR), which is close to the accretion disk and contains gas clouds that move near Keplerian velocities $>$ 1,000 km s$^{-1}$ (e.g., \cite{1996ApJ...463..144D, 2003A&A...407..461K, 2011A&A...525L...8C}). Above 1,000 km s$^{-1}$, BLs are free from contamination by star formation or supernova emission, and serve as definitive evidence for AGN activity. Moreover, the outer edge of the BLR  terminates at the hot dusty torus of the AGN, within 0.01 - 1 parsec of the SMBH (e.g., \cite{1993ApJ...404L..51N}). 

Furthermore, depending on the orientation of the AGN and the properties of the emission lines observed for each, an AGN can be classified as Type I or Type II. Type I are oriented pole-on and are observed to have BLs. On the other hand, Type II are oriented edge-on and are observed to have narrow emission lines (NLs; FWHM $\sim$ 300 - 1,000 km s$^{-1}$) in a lower density region, termed the narrow line region (NLR). The NLR is just beyond the dusty torus, out to several kiloparsecs from the SMBH (see the Unified Model of AGNs; \cite{1993ARA&A..31..473A, 1995PASP..107..803U}). However, in the NLR, star formation can produce strong narrow emission lines that mimic AGN emission (e.g., \cite{2016A&A...590A..37S}). As a result, classifying Type I AGNs using BLs, including the broad H$\alpha$ and broad H$\beta$ lines, for example, is typically more reliable than narrow lines at identifying AGNs (e.g., \cite{2015ApJS..219....1O, 2019ApJS..243...21L}). 

With the rise of the Sloan Digital Sky Survey (SDSS; \cite{ 2000AJ....120.1579Y, 2011AJ....142...72E, 2017AJ....154...28B}), an abundance of galaxy spectra have been made available. With this large population of spectra, multiple BL AGN catalogs have emerged. For example, \cite{2019ApJS..243...21L} report a sample of 14,584 BL AGN at \textit{z} $< 0.35$, which are detected from the SDSS's Seventh Data Release (DR7). The authors predominately used broad H$\alpha$ emission in their study and computed H$\alpha$ luminosities between 10$^{38.5}$ - 10$^{44.3}$ erg s$^{-1}$, FWHMs between 500 - 34,000 km s$^{-1}$, and virial SMBH masses, estimated from the BL measurements, between 10$^{5.1}$ - 10$^{10.3}$ M$_{\odot}$. Further, \cite{2010AJ....139.2360S} compiled a catalog of 105,783 unique quasars from DR7. These quasars were spectroscopically confirmed, luminous (M$_{i} < 22.0$), and featured at least one line with FWHM $>$ 1,000 km s$^{-1}$. \cite{2015ApJS..219....1O} also discovered 1,835 BL AGNs in SDSS's DR7 at $\textit{z} < 0.2$. They measured broad H$\alpha$ emission in these galaxies (FWHM $>$ 800 km s$^{-1}$ for their sample) and used observations from the $\textit{Chandra X-Ray Observatory}$ (\cite{2010ApJS..189...37E}) to confirm the AGNs. Finally, \cite{2018A&A...613A..51P} assembled a BL AGN catalog of 21,877 galaxies using the extended Baryon Oscillation Spectroscopic Survey (eBOSS), from the SDSS's fifteenth data release in its fourth phase (SDSS-IV; \cite{2017AJ....154...28B}). 

In addition, existing galaxy formation models suggest that galaxies evolve through frequent mergers with companion galaxies. (e.g., \cite{2003ApJ...582..559V}). During these mergers, the two SMBHs at the galaxy centers lose energy via dynamical friction. In some cases, one or both of the SMBHs may actively be accreting matter as an AGN. If both are AGNs, then the pair is termed ``dual AGN" (\cite{2007ApJ...660L..23G, 2009ApJ...698..956C, 2015ApJ...806..219C}). A small quantity of dual AGNs are presently known. On the other hand, if one of the galaxies is an AGN and the other quiescent, the AGN is termed an ``offset AGN'' (\cite{2014ApJ...789..112C}). One way to detect an AGN in a merging system is to use BL detections. For example, if we detect BLs outside of the nuclear region of a target galaxy, and within the nuclear region of a companion galaxy, it is an indication that the companion hosts an AGN (i.e., this is an offset AGN detection). 


In this paper, we add to the existing population of SDSS AGNs using the SDSS-IV Mapping Nearby Galaxies at Apache Point Observatory (MaNGA) integral field unit (IFU) catalog (\cite{2015ApJ...798....7B}). MaNGA is a large-scale spectroscopic survey that has observed 10,010 nearby galaxies. With our pipeline, we detect 301 broad H$\alpha$-emitting galaxies and 801 broad H$\beta$ emitting galaxies at $\ge 5\sigma$ above the continuum, with line widths $>$ 1,000 km $s^{-1}$ and BL emission in at least 1 spaxel. In total, we detect 1,042 unique BL galaxies; 60 BL galaxies feature both H$\alpha$ and H$\beta$ emission. We also report the physical parameters of these BL galaxies (e.g., BL luminosities, BL FWHMs, and SMBH masses). 

Overall, we identify 987 additional MaNGA BL AGNs compared to \cite{2015ApJS..219....1O}, for example. For reference, the BL AGNs from \cite{2015ApJS..219....1O} were detected using SDSS DR7, which observed galaxies with small (3" diameter) optical fibers that only traced a small region close to the galactic center - potentially missing nuclear activity outside of this region (e.g., \cite{2016AJ....152..197Y}). On the other hand, MaNGA can detect spatially extended galactic features, which can reveal off-nuclear activity and large-scale emission-line regions, leading to the detection of new BL AGNs. 

Moreover, we detect 35 broad H$\alpha$ galaxies and 77 broad H$\beta$ galaxies with emission offset from the galaxy center. We find that 3 of these galaxies feature both broad H$\alpha$ and H$\beta$ emission, leading to 109 unique galaxies with BL emission offset from the galaxy center in our sample.


This paper is outlined as follows: Section \ref{sec:obsbl} covers the details of the SDSS-IV MaNGA survey and its data pipeline, Section \ref{sec:databl} reviews the techniques we use to build the BL catalog and analyze the physical properties of the BLR, Section \ref{sec:resultsbl} overviews our results, and Section \ref{sec:summarybl} provides our conclusions and intended future work. All wavelengths are provided in vacuum and we assume a $\Lambda$CDM cosmology with the following values: $\Omega_{M} = 0.287$, $\Omega_{\Lambda} = 0.713$ and $H_{0} = 69.3$ $ \rm{km}$ $\rm{s}^{-1}$ $\rm{Mpc}^{-1}$. 
\section{Galaxy Observations} \label{sec:obsbl}
\subsection{MaNGA Catalog}
To assemble a statistically significant sample of spatially-resolved observations of BL galaxies, we utilize the largest IFU spectroscopic survey of galaxies to date, the SDSS-IV MaNGA catalog (\cite{2015ApJ...798....7B}). MaNGA uses the SDSS 2.5 m telescope (\cite{2006AJ....131.2332G}), and has observed the spectra for 10,010 nearby galaxies (0.01 $<$ \textit{z} $<$ 0.15; average \textit{z} $\approx$ 0.03) with stellar mass distributions between $10^{9}$ $M_{\odot}$ and $10^{12}$ $M_{\odot}$. The wavelength range of MaNGA spans 3622 \AA$\>$ - 10354 \AA, with a typical spectral resolution of $\sim$ 2000 and a velocity resolution $\sigma =$ 72 km s$^{-1}$ (\cite{2015ApJ...798....7B, 2016AJ....152...83L}). 

MaNGA uses IFU fiber bundles grouped into hexagonal grids with field-of-view (FoV) diameters between 12.$^{\prime\prime}$5 to 32.$^{\prime\prime}$5, where the size of the bundles corresponds to the apparent size of the target galaxy (\cite{2015ApJ...798....7B}). Specifically, the system is comprised of two 19-fiber IFUs (12.$^{\prime\prime}$5 FoV), four 37-fiber IFUs (17.$^{\prime\prime}$5 FoV), four 61-fiber IFUs (22.$^{\prime\prime}$5 FoV), two 91-fiber IFUs (27.$^{\prime\prime}$5 FoV), and five 127-fiber IFUs (32.$^{\prime\prime}$5 FoV). The MaNGA observations generate spectroscopic maps out to at least 1.5 times the effective radius, with an average footprint of $\sim$ 500 arcsec$^{2}$ per IFU; the typical galaxy is mapped out to a radius of 15 kpc. Each MaNGA spatial pixel, or spaxel, covers 0.$^{\prime\prime}$5 $\times$ 0.$^{\prime\prime}$5 and the average FWHM of the on-sky point spread function (PSF) is 2.$^{\prime\prime}$54, which corresponds to a typical spatial resolution of 1 -2  kpc (\cite{2015AJ....149...77D}).  


\subsection{MaNGA Data Reduction Pipeline}
The MaNGA Data Reduction Pipeline (DRP) produces sky-subtracted spectrophotometrically calibrated spectra in a FITS file format that is used for scientific analysis (\cite{2016AJ....152...83L}). The resulting DRP data product is run through MaNGA's Data Analysis Pipeline (DAP; \cite{2019AJ....158..231W}) that provides three-dimensional data cubes that combine dithered observations. The data cubes offer science data products, such as stellar kinematics, emission-line parameters (e.g., fluxes and equivalent widths), and spectral indices (e.g., D4000 and Lick indices). The data products are publicly released periodically as MaNGA Product Launches (MPLs).  To construct our catalog of BL galaxies in MaNGA, we use MaNGA's eleventh, and final, data release (MPL-11), which contains data for 10,010 unique galaxies. 
\section{Analysis} \label{sec:databl}
\subsection{Stellar Continuum Fit and Subtraction and Spectral Fitting}
The MaNGA DAP provides fits for prominent emission lines (e.g., the narrow components of the H$\alpha$, H$\beta$, and [OIII] $\lambda5007$ lines). However, it does not offer fits for the broad H$\alpha$ and H$\beta$ components (FWHM $>$ 1,000 km s$^{-1}$). As a result, we create a custom pipeline to scan for broad H$\alpha$ and  H$\beta$ emission in MaNGA.

The initial step in our pipeline is accessing the DRP to retrieve the data cubes for each MaNGA galaxy. The data cubes provide a spectrum for each spaxel across the FoV of each galaxy (spaxel arrays vary between 32 $\times$ 32 spaxels to 72 $\times$ 72 spaxels, depending on IFU configuration). We then use the spectroscopic redshifts of the stellar component of each galaxy, provided by the DAP (\texttt{STELLAR\_Z} parameter), to shift the spectra to rest vacuum wavelengths ($\lambda_{\rm{rest}} =$ 4862 \AA $\>$ for H$\beta$ and $\lambda_{\rm{rest}} =$ 6562 \AA $\>$ for H$\alpha$. For all target galaxy redshifts (0.01 $<$ \textit{z} $<$ 0.15), H$\alpha$ and H$\beta$ fall within the MaNGA wavelength range. Then, to subtract the stellar continuum for each BL galaxy’s corrected spectra, we use the software package \texttt{pPXF} (\cite{2012ascl.soft10002C,2017MNRAS.466..798C,2023MNRAS.526.3273C}). \texttt{pPXF} performs a polynomial fit on each galaxy’s spectrum, while masking prominent emission and absorption features. To fit the stellar population synthesis model on each fit, we use the MILES\footnote{\href{http://miles.iac.es/}{http://miles.iac.es/}} stellar templates library. This library features $\approx$ 1,000 stars, with spectra obtained by the \textit{Isaac Newton Telescope}, and covers the wavelength range of 3525 \AA $\>$- 7500 \AA $\>$at a 2.5 \AA $\>$FWHM resolution. We then subtract the stellar continuum.

We then apply a mask to each datacube, such that the imported wavelength range for each spectrum matches the wavelength range of the stellar templates library (3525 \AA - 7500 \AA). Next, we normalize each spectrum by dividing fluxes in this wavelength range by each spectrum's median flux value (to avoid numerical issues; similar to \cite{2023ApJ...945..127N}; see \cite{2017MNRAS.466..798C} for a more detailed discussion). Subsequently, we define a typical instrument resolution of $\approx$ 2.5 \AA, construct a set of Gaussian emission line templates (to mask emission lines; provided by \texttt{pPXF}), and fit the stellar templates. 

Then, we use the spectroscopic analysis \texttt{Python} package \texttt{PySpecKit} (\cite{2011ascl.soft09001G, 2022AJ....163..291G}) to perform a multi-Gaussian scan for the broad H$\alpha$ and broad H$\beta$ emission-line components at $\ge 5 \sigma$ above the background continuum, with FWHMs $>$ 1,000 km s$^{-1}$, in the 10,010 galaxies in MPL-11. Generally, for catalogs of Type I BL AGN, a standard line width of 1,000 km s$^{-1} $ is used to differentiate the BLR from the NLR  (e.g., \cite{2005AJ....129.1783H, 2010ApJS..189...37E}). However, it is important to note that low-luminosity AGNs and low-mass SMBHs can feature BL widths below this threshold. As referenced in \cite{2019ApJS..243...21L}, some low-mass SMBHs, which have been confirmed to be AGNs using X-ray observations (e.g., \cite{2007ApJ...670...92G, 2012ApJ...755..167D, 2018ApJS..235...40L}), can have widths as low as 500 km s$^{-1}$. In this paper, we only consider AGNs with a high enough luminosity and SMBH mass to place their line widths above 1,000 km s$^{-1}$. We acknowledge that we may miss weaker AGN and lower mass SMBHs below this threshold. 

For the H$\alpha$ fits, we then perform a multi-component fit to the NL and BL components of the H$\alpha$ line. We proceed to analyze the BL fit on the H$\alpha$ line, which has been isolated by subtracting the continuum. Note, \texttt{pPXF} factors in the stellar templates, as well as the individual BL and NL components to accurately model the stellar continuum, and the presence of a BL does not affect the quality of the continuum modeling, compared to a single component fit (see \cite{2012ascl.soft10002C,2017MNRAS.466..798C,2023MNRAS.526.3273C}). In addition, in some instances, it can be challenging to accurately measure the width of the broad H$\alpha$ line, especially near the strong neighboring emission lines of [NII] $\lambda$$\lambda$6549,6585. Due to the complexities of this, we follow the procedure performed by \cite{2015ApJS..219....1O}. In particular, we calculate a ratio between the mean fluxes of two spectral regions near H$\alpha$: 6460 - 6480 \AA $\>$ and 6523 - 6543 \AA. The former is near H$\alpha$ but not too close to be mixed with H$\alpha$ emission, which makes it a good proxy for the continuum near H$\alpha$.  The latter region is closer to the H$\alpha$ line and is a good tracer of the flux of the broad component of the H$\alpha$ line. Once this ratio is computed for each fit, we require that this value be $>$ 1 for all MaNGA galaxies, and pass our other requirements, to be considered a broad H$\alpha$ detection. 

\begin{figure}
	\includegraphics[width = 8.5cm]{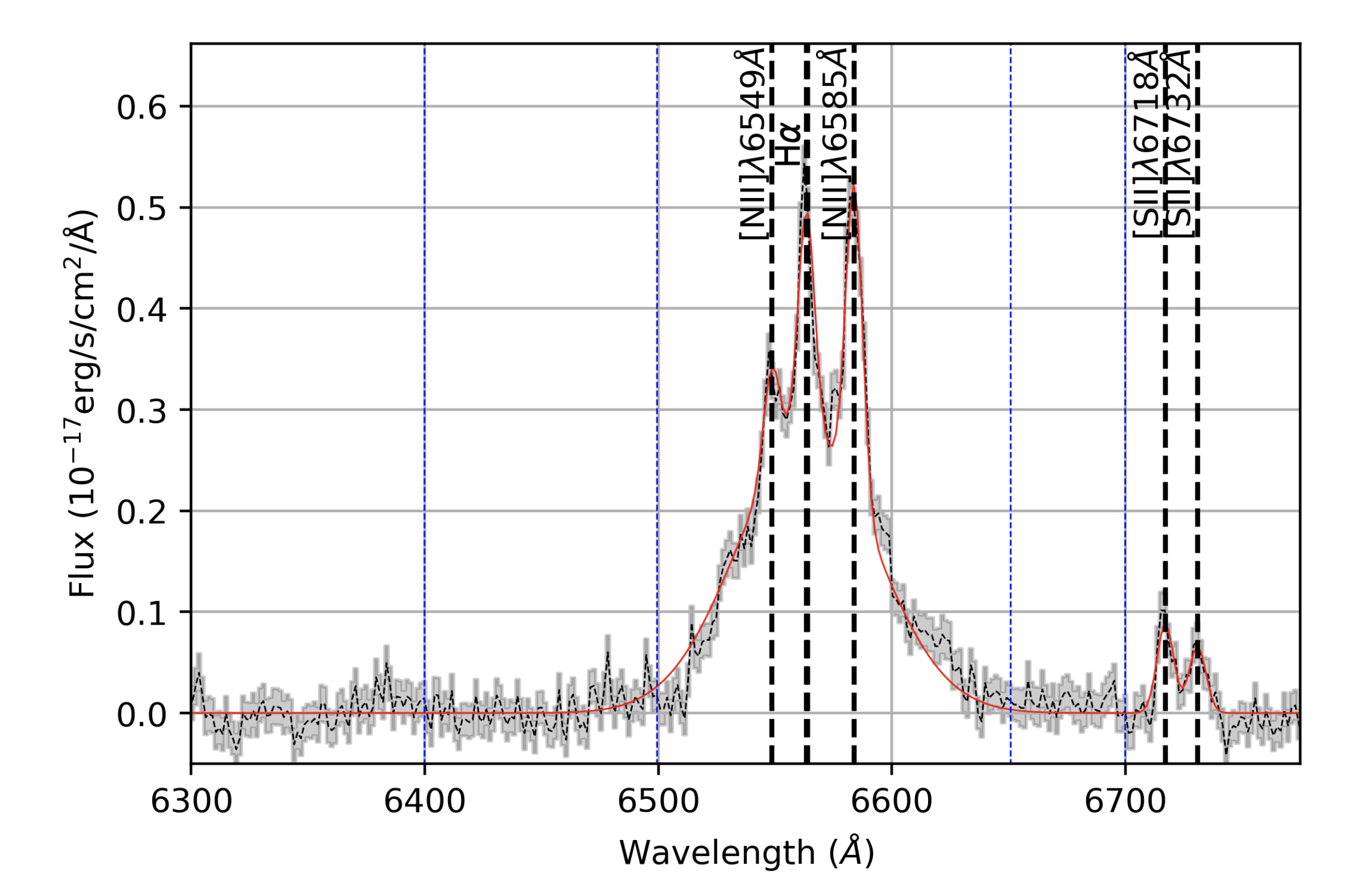}
	\caption{A sample spectrum from an individual spaxel showing the broad H$\alpha$ line detected at $\ge 5 \sigma$ above the continuum in J004730.34+154149.4. The dotted black line is the stellar continuum subtracted spectrum, the shaded gray region is the uncertainty, the solid red line represents the best fit, the bold dashed black lines signify the rest wavelengths of the H$\alpha$, $\rm{[}$NII$\rm{]}$ $\lambda\lambda$6549, 6585, and $\rm{[}$SII$\rm{]}$  $\lambda\lambda$6718, 6732 lines, and the two sets of blue dotted vertical lines correspond to the neighboring continuum windows where the RMS flux values of the continuum are calculated.}
	\label{fig:blspectra}
\end{figure}

\begin{figure}[t]
	\includegraphics[width = 8.5cm]{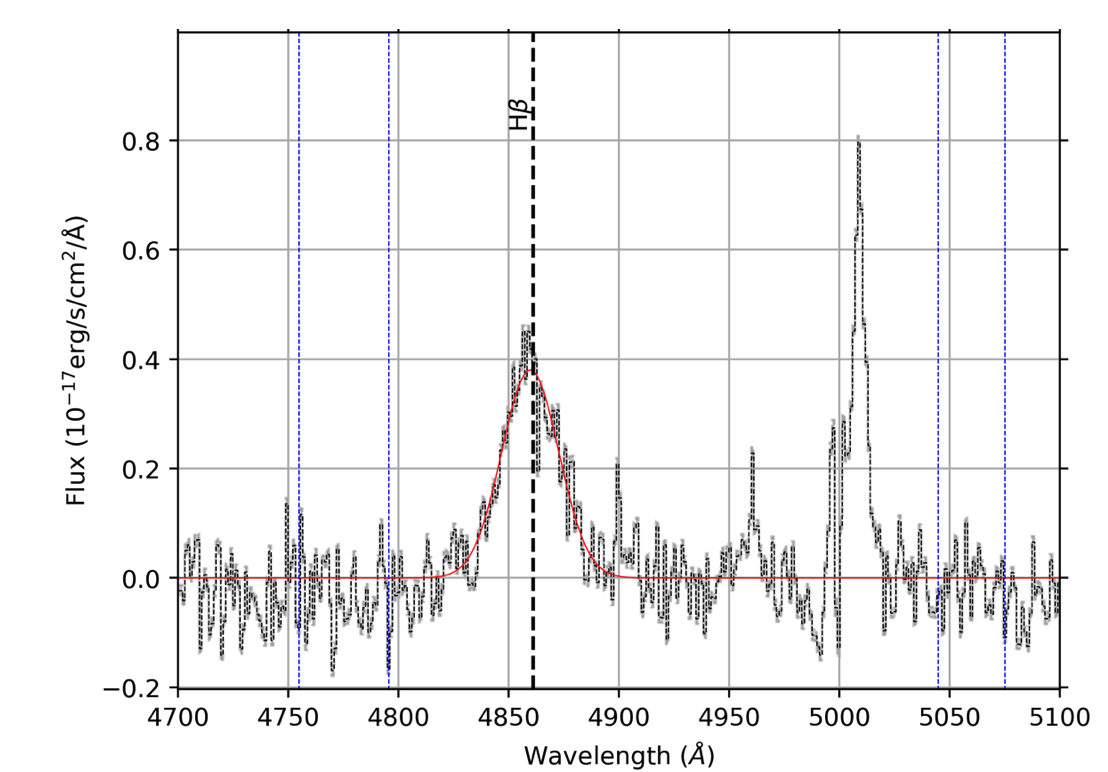}
	\caption{A sample spectrum from an individual spaxel showing the broad H$\beta$ line detected at $\ge 5 \sigma$ above the continuum in J171411.63+575834.0. The dotted black line is the stellar continuum subtracted spectrum, the shaded gray region is the uncertainty, the solid red line represents the best fit, the bold dashed black line signifies the rest wavelength of the H$\beta$ line, and the two sets of blue  dotted vertical lines correspond to the neighboring continuum windows where the RMS flux values of the continuum are calculated}
	\label{fig:blspectra2}
\end{figure}

For the H$\beta$ line, we fit NL and BL components to the line profile and analyze the BL results. We use the conservative 5$\sigma$ and 1,000 km s$^{-1}$ thresholds to ensure we identify definitive BL emission, and only require one spaxel to have a BL detection to count as a BL galaxy. Figures \ref{fig:blspectra} and \ref{fig:blspectra2} show examples of broad H$\alpha$ and H$\beta$ fits from our pipeline.


We measure broad H$\alpha$ FWHMs between 1,010 km s$^{-1}$ - 4,919 km s$^{-1}$, with a median FWHM of 2,079 km s$^{-1}$. For the broad H$\beta$ galaxies, we calculate broad H$\beta$ FWHMs between 1,001 km s$^{-1}$ - 5,849 km s$^{-1}$, with a median FWHM of 1,146 km s$^{-1}$. Doing so, we avoid potential contamination by star-formation, which can produce strong narrow emission lines $<$ 1,000 km s$^{-1}$ (e.g., \cite{2016A&A...590A..37S}).


\subsection {Broad Line Flux/ Velocity Maps and Line Luminosities}
\label{blflux}
\begin{figure*}
            \centering
		\includegraphics[width = 17.5cm]{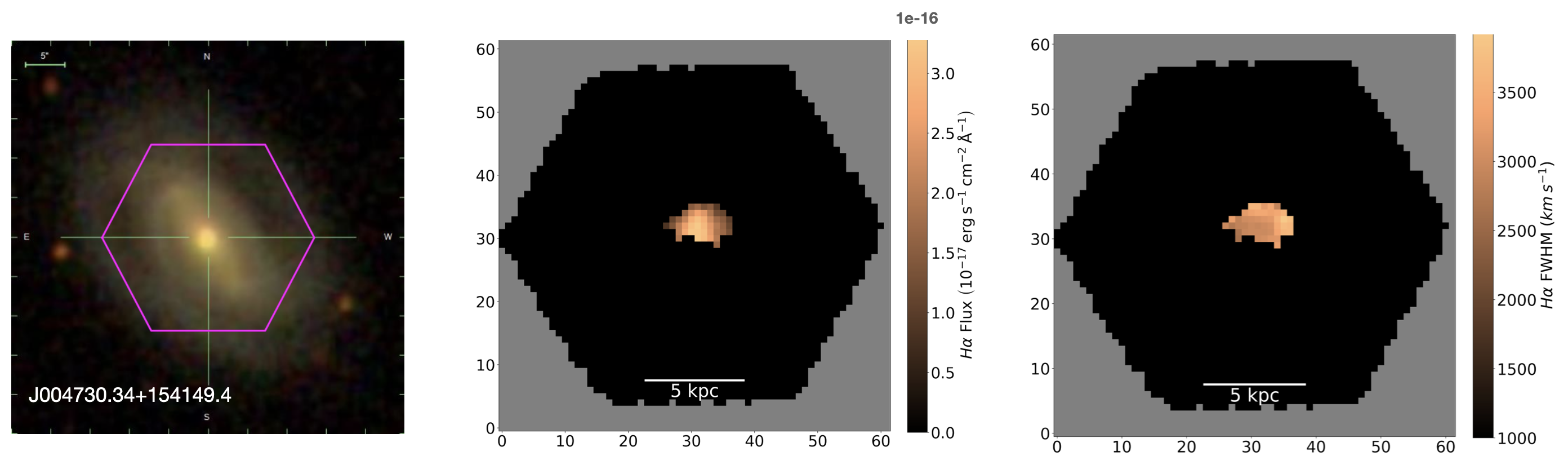}
	\includegraphics[width = 17.5cm]{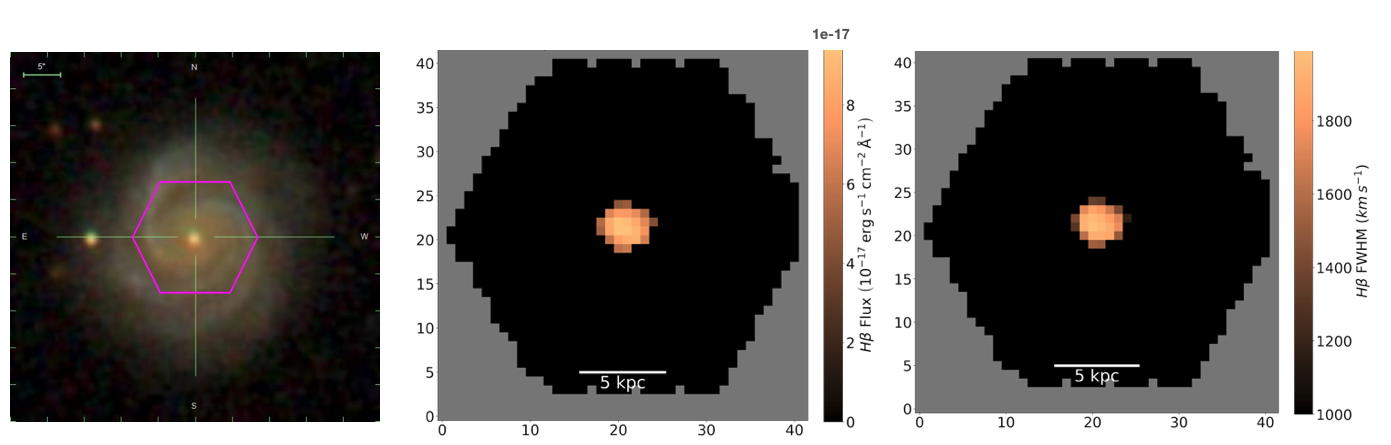}
	\caption{[Top] From left to right, an example SDSS optical image from MaNGA, a custom generated broad H$\alpha$ flux map, and a custom generated broad H$\alpha$ FWHM map for J004730.34+154149.4. [Bottom] From left to right, an example SDSS optical image from MaNGA, a custom generated broad H$\beta$ flux map, and a custom generated broad H$\beta$ FWHM map for J212851.19-010412.4. The gray region is outside of the MaNGA FoV and the black region shows spaxels with no BL emission. North is up and east is to the left.}
	\label{fig:blmaps}
\end{figure*}

We create custom BL ﬂux and velocity maps to analyze the strength and distribution of the BLs, and their cloud velocities, in our sample. We create these maps using the integrated BL ﬂux value and measured FWHM from each spectrum for each spaxel for each BL galaxy. We measure these values during the spectral fitting routine. Examples of broad H$\alpha$ and H$\beta$ flux and velocity maps are shown in Figure \ref{fig:blmaps}. Note, these BL flux maps trace the presence of an AGN, oriented so that the central SMBH is viewed directly.

 Moreover, evaluating the AGN luminosities for our sample helps us determine the range of broad H$\alpha$ and broad H$\beta$ luminosities for the BL galaxies in MaNGA above our FWHM cutoff of 1,000 km s$^{-1}$. 
	
	We determine the line luminosity of the broad component for each spaxel using Equations \ref{eqn:ha_lum} and \ref{eqn:hb_lum}:
	\begin{equation}
		\label{eqn:ha_lum}
		L_{H\alpha} =  F_{H\alpha} (4\pi R^{2})
	\end{equation}
	\begin{equation}
		\label{eqn:hb_lum}
		L_{H\beta} =  F_{H\beta} (4\pi R^{2})
	\end{equation}
	where F$_{H\alpha}$ is the measured H$\alpha$ flux, summed across all BL spaxels from our spectral fits, F$_{H\beta}$ is the measured H$\beta$ flux, summed across all BL spaxels from our spectral fits, and R is the DAP-provided luminosity distance based on the galaxy redshift.  
 
	\subsection {SMBH Masses}
	\label{lum}

 Virial SMBH masses for low-redshift AGNs can typically be estimated using measurements of the optical continuum strength (e.g., the luminosity of the continuum at 5100 \AA) and the width of the broad H$\beta$ line (e.g., \cite{2005ApJ...630..122G}). However, systematic uncertainties and difficulties in measuring these quantities can make this method challenging. Additionally, if the optical continuum of radio-loud AGNs is enhanced by emission from AGN jets, the continuum, and SMBH masses will be systematically over-estimated (e.g., \cite{2005ApJ...630..122G}). Therefore, \cite{2005ApJ...630..122G} analyzed a sample of SDSS BL AGNs and proposed a new formalism for calculating SMBH masses using BL measurements. In their sample of $\sim$ 3,000 AGNs, they found that: H$\alpha$ luminosity scales nearly linearly with the optical continuum luminosity, there is a definitive correlation between H$\alpha$ and H$\beta$ line widths (i.e. FWHMs), and that SMBH masses can be estimated solely using observations of the broad H$\alpha$ emission line. They also uncovered that the H$\beta$ emission line can also be solely used if H$\alpha$ emission is not available. These mass measurements, for example, are critical in the larger context of resolving the role of SMBHs in their host galaxy's evolution (e.g., the M-$\sigma$ relationship, which is the empirical correlation between a galaxy's SMBH mass and the stellar velocity dispersion of its galactic bulge; \cite{2000ApJ...539L..13G, 2000ApJ...539L...9F}). 
 
 Here, we use the SMBH mass estimator from \cite{2005ApJ...630..122G}. The SMBH mass is derived using the broad H$\alpha$ and broad H$\beta$ luminosities in Equations \ref{eqn:bh_mass} and \ref{eqn:bh_mass2}:
	\begin{equation}
		\begin{aligned}
		\label{eqn:bh_mass}
		M_{\rm{BH}} = (1.3 \pm 0.3) \times 10^{6} \Big(\frac{L_{H\alpha}}{10^{42}\rm{erg}\rm{s}^{-1}}\Big)^{0.57\pm0.06}\\ \Big(\frac{\rm{FWHM}_{H\alpha}}{10^{3} \rm{km}\rm{s}^{-1}}\Big)^{2.06\pm 0.06}M_{\odot}
		\end{aligned}
	\end{equation}
	\begin{equation}
		\begin{aligned}
		\label{eqn:bh_mass2}
		M_{\rm{BH}} = (3.6 \pm 0.2) \times 10^{6} \Big(\frac{L_{H\beta}}{10^{42}\rm{erg}\rm{s}^{-1}}\Big)^{0.56\pm0.02}\\ \Big(\frac{\rm{FWHM}_{H\beta}}{10^{3} \rm{km}\rm{s}^{-1}}\Big)^{2}M_{\odot}
		\end{aligned}
	\end{equation}
	where L$_{H\alpha}$ is the luminosity of the broad H$\alpha$ component, summed across all BL spaxels from our spectral fits, L$_{H\beta}$ is the luminosity of the broad H$\beta$ component, summed across all BL spaxels from our spectral fits, FWHM$_{H\alpha}$ is the median broad H$\alpha$ FWHM value measured across all BL spaxels for each BL galaxy, and FWHM$_{H\beta}$ is the median broad H$\beta$ FWHM value measured across all BL spaxels  for each BL galaxy. 
  
	\subsection{H$\alpha$ and H$\beta$ Companion Galaxies}
	\label{sec:blmergers}
	SMBHs are suspected to form and grow through two primary paths: 1) active accretion, whereby nearby gas accumulates onto the central engine, and 2) hierarchical merging of separate SMBHs through large-scale mergers (e.g., \cite{2013ARA&A..51..511K}). As detailed in, e.g., \cite{2021ApJ...914...37B}, the byproduct of mergers is the formation of a gravitationally bound SMBH system, where each galaxy's SMBH moves towards the center of the system due to dynamical friction (\cite{1943ApJ....97..255C}). For some mergers, there may be an offset or dual AGN, where the former signifies one galaxy in a merging system hosts an AGN, and the latter signifies that both do. If offset AGNs exist in merging systems in our sample, we expect to find BL emission in either the target or companion galaxy. Note, we search for offset AGNs by identifying galaxies with BL emission solely beyond the center of the target MaNGA  galaxy, which we approximate to be 2.$^{\prime\prime}$5, or $\sim$ 1-2 kpc from the target galaxy's center. Further, we anticipate a slight velocity offset for the companion, as it is likely at a slightly different redshift than the target galaxy. On the other hand, if dual AGN are in merging systems in our sample, we expect to find BL emission both in the target galaxy and the companion, or in two companion galaxies if we detect a triple merging system.

	Finally, to further determine if a possible connection exists between BL emission and the presence of a companion galaxy, we consider if a BL galaxy is merging or not based on the catalog produced by \cite{2023MNRAS.522....1N} (``Nevin catalog" hereafter). The authors calculate the merger probability for the 1.3 million galaxies in the SDSS DR16 photometric sample, using imaging predictors that have been trained to separate mock images of simulated merging and non-merging galaxies  (see \cite{2019ApJ...872...76N}). We analyze the BL galaxies in the SDSS DR16 Nevin catalog and classify a BL galaxy as a merger if the Nevin catalog gives it a merger probability (p$_{\rm{merg}}$) $>$ 0.5. 
	\section{Results}
	\label{sec:resultsbl}
	In this section, we present our findings for the BL galaxies in MaNGA. In total, we find 301 galaxies with broad H$\alpha$ emission at $\ge$ 5$\sigma$ above the background continuum, with FWHM values $>$ 1,000 km s$^{-1}$, in MaNGA’s MPL-11; 801 galaxies with broad H$\beta$ emission. In total, we detect 1,042 unique BL galaxies; 60 feature both H$\alpha$ and H$\beta$ emission ( Appendix \ref{appendixfff}). While H$\alpha$ is the strongest line in the Balmer series and is one of the most reliable signatures of a Type I AGN, we note that the multifaceted  H$\alpha$-[NII] complex, which produces some blending of the H$\alpha$ and [NII] line profiles and can create asymmetric broad H$\alpha$ profiles, which may lead to some missed broad H$\alpha$ detections. This may account for the discrepancy in the number of broad H$\alpha$ (301) vs. H$\beta$ (801) galaxies we detect. 	
 
    For the broad H$\alpha$ galaxies, we detect 35 galaxies, where the BL emission is spatially offset from the center of the galaxy ($\sim$ 12\% of the sample); 77 for the H$\beta$ galaxies ($\sim$ 10\%). Note, 3 galaxies with offset BL emission feature both broad H$\alpha$ and H$\beta$ emission, leading to 109 unique galaxies with BL emission offset from the center of the galaxy. We also measure the broad H$\alpha$ and broad H$\beta$ luminosities for our sample to range between 10$^{37}$ - 10$^{43}$ erg s$^{-1}$.
	
	\subsection{Radius of the BLR}
    \label{blr_sec}
\begin{figure}
		\includegraphics[width = 8.5cm]{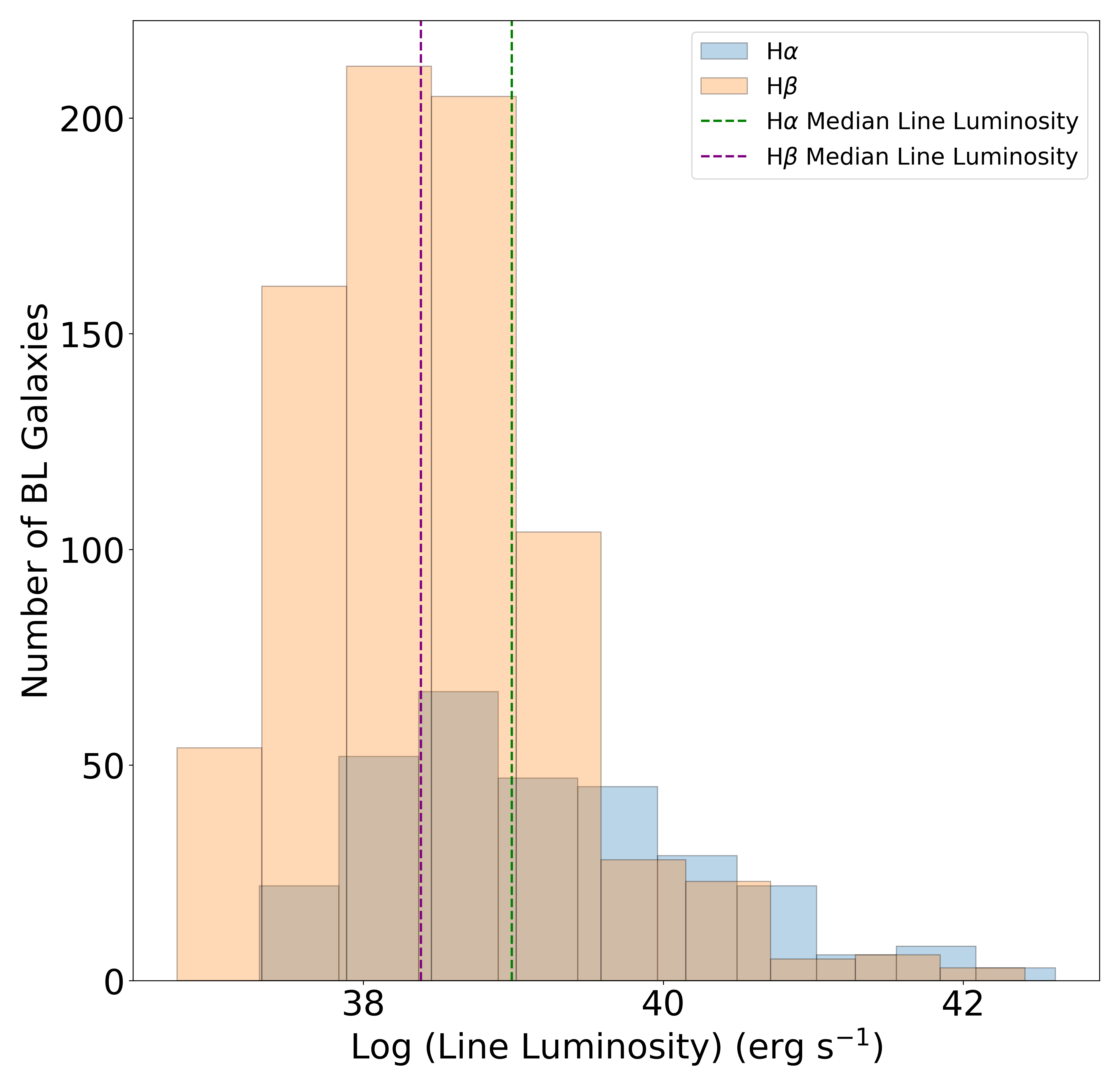}
		\caption{BL Luminosity distribution for the BL galaxies. The blue histogram represents the 301 broad H$\alpha$ galaxies and the orange histogram represents the 801 broad H$\beta$ galaxies. Line luminosities for galaxies with both broad H$\alpha$ and broad H$\beta$ emission are shown for each line independently. The dotted green line marks the median luminosity (9.7 $\times$ 10$^{38}$ erg s$^{-1}$) for the H$\alpha$ sample, whereas the purple dotted line marks the median luminosity (2.4 $\times$ 10$^{38}$ erg s$^{-1}$) for the H$\beta$ sample.}
		\label{fig:bllum}
		
	\end{figure}

	The line intensity and line widths of each BL can help trace the features of an AGN (e.g., the strength of the AGN and the BLR radius). For our sample of BL galaxies, we display the broad H$\alpha$ and broad H$\beta$ luminosities in Figure \ref{fig:bllum}. We determine that the broad H$\alpha$ galaxies are more luminous, with a median luminosity of  9.7 $\times$ $10^{38}$ erg s$^{-1}$, which is a factor of 4 times larger than the median luminosity, 2.4 $\times$ 10$^{38}$ erg s$^{-1}$, of the H$\beta$ galaxies. Previous studies have suggested that the profiles of the Balmer lines can vary from one another, indicative of higher density and higher velocity zones of the BLR (e.g., \cite{1982ApJ...259...48S, 1984ApJ...280..491S, 1986ApJS...62..821C, 1989A&A...211..293V}). Additionally, some authors report that H$\beta$ emission is much fainter than H$\alpha$ ($\sim$ $\frac{1}{3}$ in the absence of extinction; e.g., \cite{2010A&A...519A..40A}), which can be traced to the transition probabilities for the Balmer series lines, including H$\alpha$ and H$\beta$. Specifically, while the luminosities of the lines vary depending on the physical conditions of the gas emitting these lines (e.g., temperature, density, and ionization state), in general, the probability of a hydrogen atom transitioning to the excited state corresponding to the H$\alpha$ line (electron transitions from n = 3 energy level to n = 2 energy level) is higher than that for the H$\beta$ line (electron transitions from n = 4 energy level to n = 2 energy level; i.e., less energy involved; e.g., \cite{2006agna.book.....O}). As a result, H$\alpha$ emission is usually stronger and more luminous, and we contend that this is a viable explanation for why the broad H$\alpha$ luminosities are larger than the broad H$\beta$ luminosities in our sample. 

 Furthermore, we consider the spatial extent of the BLR. As outlined in \cite{2015ApJ...798....7B}, the PSF of MaNGA is expected to have a FWHM of 2.$^{\prime\prime}$5, or 5 spaxels, which corresponds to a 1-2 kpc resolution. However, as reviewed in \cite{2021ApJS..257...66C}, MaNGA IFU observations are liable to spatial information degradation. This is due to a combination of atmospheric seeing, defects in the optics and telescope hardware, and sampling size characteristics. The authors also acknowledge that physical gaps between sampling elements can enlarge the effective PSF size. As a result, IFU data are typically smoothed and processed to become spatially correlated. To address this, many efforts to reduce the limitations of the PSF of IFU data have been conducted, which include forward modeling, deconvolution algorithms, and optimal spectrum extraction (e.g., \cite{2000ApJ...529.1136C, 2003AJ....125.2266L, 2008MNRAS.390...71C, 2015AJ....150...92B, 2021ApJS..257...66C}). In light of these complexities, we elect to measure the size of the BLR, for the broad H$\beta$ galaxies, using the broad H$\beta$ luminosity (Equation \ref{eqn:r_hb}). This physically derived approach provides a more reliable estimation of the physical extent of the BLR, compared to the IFU-measured physical distances provided by MaNGA. 
 
    Additionally, in low-density and low optical depth gas, H$\beta$ emission is effective at tracing the ionizing continuum luminosity (e.g., the 5100 \AA $\>$ luminosity; \cite{2005ApJ...629...61K}). Moreover, some studies show a strong dependence between the ionizing AGN flux and H$\beta$ emission in the BLR. This is attributed to the effective temperature at the disk radius, which corresponds to the location of the BLR. As inferred from the H$\beta$ line, this temperature is 1,000 Kelvin, close to the threshold that dust can exist at (see \cite{2000ApJ...533..631K,  2004A&A...424..793W, 2009ApJ...697..160B, 2011A&A...525L...8C, 2013ApJ...767..149B, 2016ApJ...827...53W}). As a result, broad H$\beta$ emission is often used to determine the size of the BLR (``Reverberation Mapping"; e.g., \cite{1997iagn.book.....P, 2004A&A...424..793W, 2005ApJ...629...61K}). The size of the BLR can thus be derived using the following:
	\begin{equation}
		\begin{aligned}
		\label{eqn:r_hb}
		\rm{Log}\>\rm{R}_{\rm{BLR}}\> (\rm{light-days}) = (1.381 \pm 0.080) \\+ (0.684 \pm 0.106)\>\rm{Log}\>\Big(\frac{L_{H\beta}}{10^{42}\>\rm{erg}\>\rm{s}^{-1}}\Big)
		\end{aligned}
	\end{equation}
	
	where R$_{\rm{BLR}}$ is the radius of the BLR and L$_{H\beta}$ is the luminosity of the broad H$\beta$ component.  We calculate R$_{\rm{BLR}}$ values between 0.01 - 46 light days for our H$\beta$ sample, with a median radius of 0.1 light days (0.02 pc). Comparatively, \cite{2004A&A...424..793W} investigated 34 AGNs with L$_{\rm{H}\beta}$ values between 10$^{39}$ - 10$^{43}$ erg s$^{-1}$. The authors report R$_{\rm{BLR}}$ values between 1.4 - 387 light days. Since the lower range of our broad H$\beta$ luminosities is 10$^{37}$ erg s$^{-1}$, two orders of magnitude lower than the lowest H$\beta$ luminosity in the \cite{2004A&A...424..793W} sample (10$^{39}$ erg s$^{-1}$), and the top range of our broad H$\beta$ luminosities is on the order of 10$^{42}$, one order of magnitude lower than the \cite{2004A&A...424..793W} sample (10$^{43}$ erg s$^{-1}$), we reason that our R$_{\rm{BLR}}$ values are consistent with \cite{2004A&A...424..793W} and are reasonable approximations.

	
	\subsection{SMBH Mass Estimates}
\begin{figure}
	\includegraphics[width = 8.5cm]{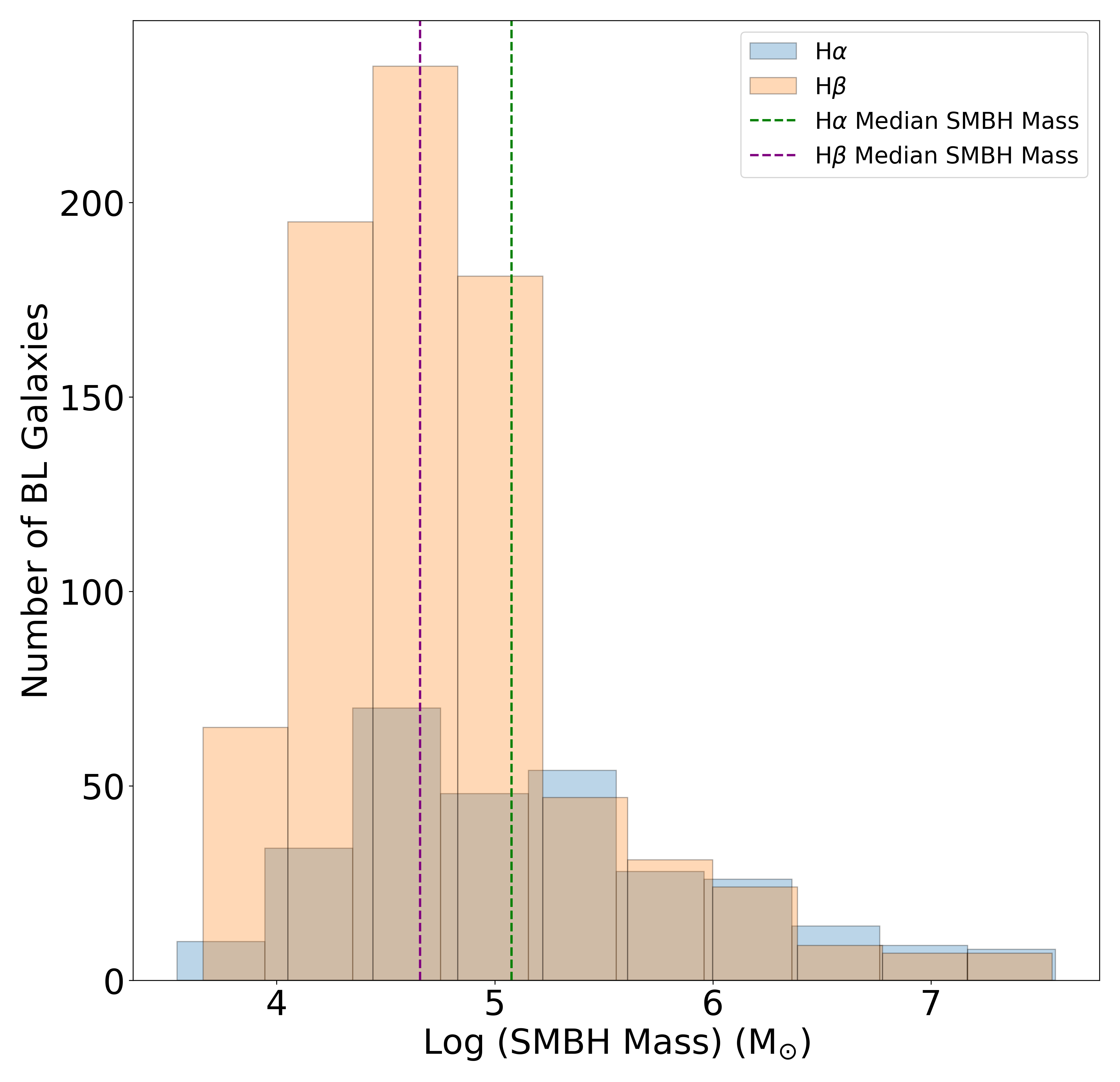}
	\caption{SMBH Mass distribution for the BL galaxies. The blue histogram represents the 301 broad H$\alpha$ galaxies (60 of which also feature broad H$\beta$ emission); the orange histogram represents 741 broad H$\beta$ galaxies (with only broad H$\beta$ emission). The dotted green line marks the median SMBH mass for the H$\alpha$ sample (\textbf{1.2 $\times$ 10$^{5}$ M$_{\odot}$}), whereas the purple dotted line marks the median SMBH mass for the H$\beta$ sample (\textbf{4.5 $\times$ 10$^{4}$ M$_{\odot}$}). For the 60 BL galaxies with broad H$\alpha$ and H$\beta$ emission, we compute SMBH mass measurements using only the broad H$\alpha$ emission-line components.}
	\label{fig:bl_mass}
	
\end{figure}

\textbf{We measure SMBH masses for the broad H$\alpha$ and H$\beta$ galaxies to range between 10$^{3}$ - 10$^{8}$ M$_{\odot}$ (Figure \ref{fig:bl_mass}), with a median mass of 1.2 $\times$ 10$^{5}$ M$_{\odot}$ for the broad H$\alpha$ galaxies and 4.5 $\times$ 10$^{4}$ M$_{\odot}$ for the broad H$\beta$ galaxies. To account for the lower end of the SMBH mass estimates (10$^{3}$ - 10$^{6}$ M$_{\odot}$), we consider the quantity of BL spaxels detected in these lower SMBH mass galaxies. In general, we find fewer BL spaxels in these galaxies, compared to the full catalog of BL galaxies. On average, we detect 7 BL spaxels in the broad H$\alpha$ galaxies with SMBH masses $<$ 10$^{6}$ M$_{\odot}$, 65 BL spaxels in the broad H$\alpha$ galaxies with SMBH masses $\ge$ 10$^{6}$ M$_{\odot}$; 7 and 81 BL spaxels for the broad H$\beta$ galaxies, respectively. Therefore, we suspect that we may only be detecting a fraction of the total BLR for the lower SMBH mass galaxies in our catalog, and that the mass estimates we provide here are lower bound estimates for them. Further, we also consider the following possibilities: 1) some of the lower SMBH masses do indeed correspond to under-massive black holes, 2) the single-epoch spectroscopic SMBH mass scaling relations from \cite{2005ApJ...630..122G} are not as accurate for the lower line luminosities found in our sample and 3) that some of the lowest luminosity BLs may feature the presence of a more complex continuum, that \texttt{pPXF} and the uniform continuum model don't account for, which may lead to under-estimates in SMBH masses.} 

Moreover, for the 60 BL galaxies with broad H$\alpha$ and H$\beta$ emission, we compute SMBH mass measurements using only the broad H$\alpha$ emission-line components (as prescribed in \cite{2005ApJ...630..122G} and consistent with SMBH mass measurements using broad H$\beta$ emission-line components). Additionally, we reason that the higher transition probability of the H$\alpha$ line, compared to the H$\beta$ line, yields larger broad H$\alpha$ luminosities, compared to broad H$\beta$ luminosities, in our sample (Section \ref{blr_sec}). As such, the measured broad H$\alpha$ SMBH masses, which scale with broad H$\alpha$ luminosities (Equations \ref{eqn:bh_mass} and \ref{eqn:bh_mass2}), are also larger than the broad H$\beta$ SMBH masses, which scale with broad H$\beta$ luminosities, in our sample.

\subsection{BL Mergers}
\label{sec:blcomp}
The BL velocity profiles may feature a blend of many different components, which include, but are not limited to, Doppler motions, inflows/ outflows (possibly induced by mergers), shocks, and/ or rotation. These velocity features can manifest to make the final profile a combination of many components  (e.g., \cite{1999ApJ...521L..95P, 2013A&A...558A..26K}). 

For our sample of BL galaxies, the BL Gaussian fits we use specifically trace the Doppler motions within the gravitational influence of the SMBH within the target galaxy, centered in MaNGA's field of view (e.g., \cite{2013A&A...549A.100K}). However, we determine that a small population of BL profiles traces the BLR in a companion galaxy. 

\begin{figure*}[h]
	\centering
	\includegraphics[width = 15cm, height = 13.5cm]{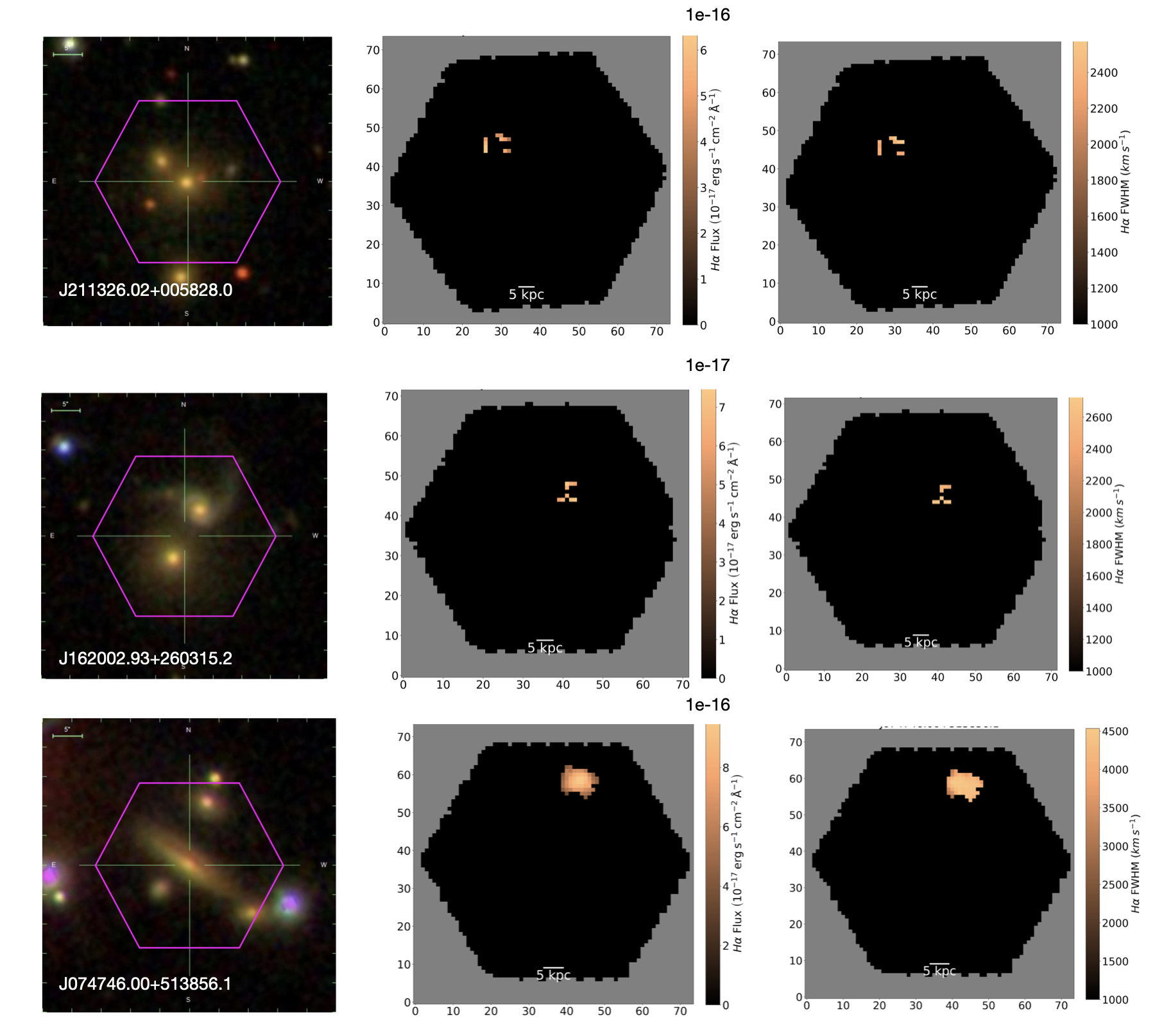}
	\caption{BL flux and velocity maps for 3/35 BL galaxies in our sample with offset broad H$\alpha$ emission. From top to bottom and left to right: SDSS optical images, broad H$\alpha$ flux maps, and broad H$\alpha$ FWHM maps for J211326.02+005828.0, J162002.93+260315.2, and J074746.00+513856.1.}
	\label{fig:blspatial}
\end{figure*}

Using the Nevin catalog (Section \ref{sec:blmergers}), we determine that 122 out of 275 broad H$\alpha$ galaxies, which we could determine a merger classification for, are currently undergoing a galaxy merger (44\%). Moreover, out of these 122 merging systems, we scan for the systems with BL emission from the companion galaxy outside of the target galaxy's nuclear region, which we define as the central 2.$^{\prime\prime}$5 region. Using these constraints, we identify 35 broad H$\alpha$ galaxies with BL emission offset from the galaxy center and present an example of three of these galaxies in Figure \ref{fig:blspatial}. 

\begin{figure*}
	\centering
	\includegraphics[width = 15cm, height = 13.5cm]{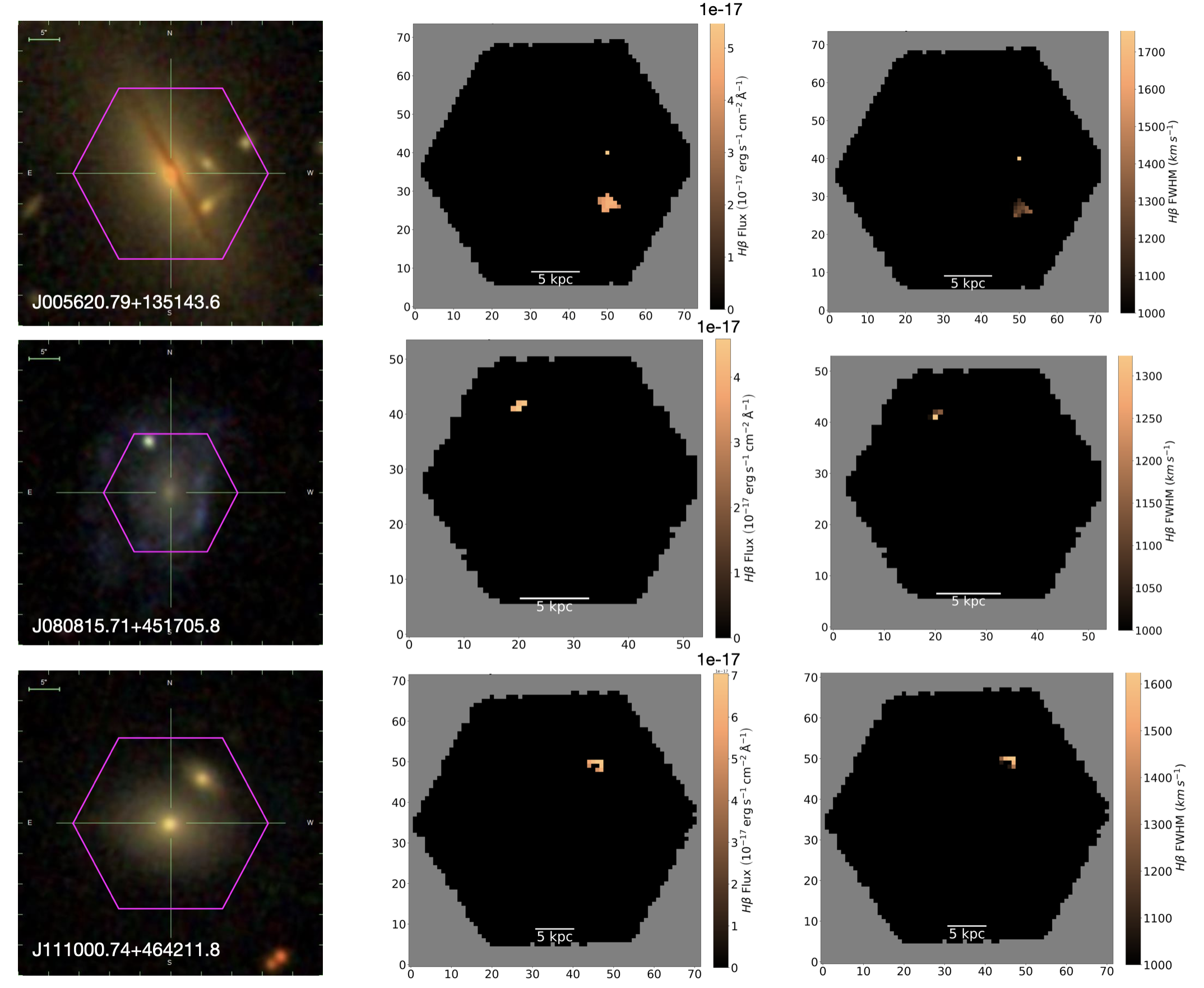}
	\caption{BL flux and velocity maps for 3/77 BL galaxies in our sample with offset broad H$\beta$ emission. From top to bottom and left to right: SDSS optical images, broad H$\beta$ flux maps, and broad H$\beta$ FWHM maps for J005620.79+135143.6, J080815.71+451705.8, and J111000.74+464211.8. In  J005620.79+135143.6, we detect broad H$\beta$ emission in two companion galaxies.}
	\label{fig:blspatial2}
\end{figure*}
Likewise, we use the Nevin catalog to detect mergers in the H$\beta$ sample. We find that 313 out of 722 broad H$\beta$ galaxies, that we could determine a merger classification for, are experiencing a galaxy merger (43\%). 

Further, we apply the same criteria as we did for the broad H$\alpha$ galaxies to search for broad H$\beta$ emission offset from the galaxy center (BL emission not within the central 2.$^{\prime\prime}$5 region). We detect 77 broad H$\beta$ that fit this criteria in our sample. We display the optical SDSS images, as well as the broad H$\beta$ fluxes and broad H$\beta$ FWHMs for 3 of these galaxies in Figure \ref{fig:blspatial2}. For J005620.79+135143.6, we identify broad H$\beta$ emission in two companion galaxies - the only definitive example in our sample. Note, 3 galaxies with offset BL emission feature both broad H$\alpha$ and H$\beta$ emission, leading to 109 unique galaxies with BL emission offset from the galaxy center in our sample.

In addition, we use the Nevin catalog to scan the total MaNGA catalog. We identify 2518 mergers, out of the 9582 MaNGA galaxies with a Nevin catalog merger classification (26\%). As a result, the broad H$\alpha$ and H$\beta$ galaxies trace a higher fraction of merging galaxies compared to the full MaNGA catalog (larger by a factor of 1.8 and 1.7, respectively). Comparatively, \cite{2023ApJ...942..107S}
identified 391 galaxy pairs within the fields of view of 8585 MaNGA IFUs and found 105 AGN in the pair sample using Baldwin-Phillips-Terlevich optical emission line diagnostic diagrams (\cite{1981PASP...93....5B, 2001ApJ...556..121K, 2003MNRAS.346.1055K}). Overall, they  report that galaxy pairs have a greater fraction of AGN than control galaxies in MaNGA, in agreement with our results. 

Moreover, our findings are also consistent with the analyses performed by \cite{2012ApJ...746L..22K, 2017ApJ...838..129B,2021ApJ...923...36S,2023ApJ...951...92B,2023AAS...24136030S}. The authors posit that during the galaxy merging process, gas and dust are driven toward the centers of the merging galaxy pairs, which triggers SMBH growth, and funnels matter to accrete onto one or both of the SMBHs, leading to AGN activation. \cite{2021ApJ...923...36S} compile a catalog of 204 offset and dual AGNs with a median \textit{z} $\sim$ 1.15 and stellar bulge separations $<$ 20 kpc, using the Advanced Camera for Surveys on the \textit{Hubble Space Telescope}. They found that AGN activation is primarily stagnant between 20 - 14 kpc, has a bump between 14 - 11 kpc, drops slightly from 11 - 4 kpc, and increases substantially from 3 - 2 kpc. We reason that the interactions between merging galaxies in our BL catalog, where the typical MaNGA footprint is mapped out to a radius of 15 kpc, may trigger some of the AGNs we detect due to the dynamic interactions between the galaxies. Ultimately, this may account for the higher merger fractions we measure our BL sample, compared to the the broader MaNGA catalog. 
\section{Summary and Future Work}
We investigate 301 broad H$\alpha$ galaxies and 801 broad H$\beta$ galaxies in MaNGA's MPL-11 (1,042 unique BL galaxies; 60 BL galaxies with both broad H$\alpha$ and broad H$\beta$ emission), which contains 10,010 unique galaxies.

We build the largest catalogs of broad H$\alpha$ and H$\beta$ galaxies in MaNGA to date, using our custom pipeline, which measures BL emission at $\ge$ 5$\sigma$ above the background continuum, and with FWHM values $>$ 1,000 km s$^{-1}$. 

Our primary findings are:
\begin{enumerate}
	\item  The radius of the BLR ranges between 0.01 - 46 light days, with a median radius of 0.1 light days (0.02 pc) for our H$\beta$ sample. 
	\item Broad H$\alpha$ and broad H$\beta$ luminosities vary between 10$^{37}$  - 10$^{43}$ erg s$^{-1}$. Further, broad H$\alpha$ galaxies are more luminous, with a median luminosity of  9.7 $\times$ $10^{38}$ erg s$^{-1}$, which is a factor of 4 times higher than the median luminosity of 2.4 $\times$ 10$^{38}$ erg s$^{-1}$ of the broad H$\beta$ galaxies. 
	\item Broad H$\alpha$ FWHMs vary between 1,010 km s$^{-1}$ - 4,919 km s$^{-1}$, with a median FWHM of 2,079 km s$^{-1}$; 1,001 km s$^{-1}$ - 5,849 km s$^{-1}$, with a median FWHM of 1,146 km s$^{-1}$ for the broad H$\beta$ galaxies.
	\textbf{\item SMBH masses for the H$\alpha$ and H$\beta$ galaxies range between 10$^{3}$ - 10$^{8}$ M$_{\odot}$, with a median mass of 1.2 $\times$ 10$^{5}$ M$_{\odot}$ for the broad H$\alpha$ galaxies and 4.5 $\times$ 10$^{4}$ M$_{\odot}$ for the broad H$\beta$ galaxies.}  
	\item 122 out of 275 broad H$\alpha$ galaxies, which we could determine a merger classification for, are currently undergoing a galaxy merger (44\%). Similarly, 313 out of 722 broad H$\beta$ galaxies, that we could determine a merger classification for, are experiencing a galaxy merger (43\%). Both are well above the merger fraction in the full MaNGA sample (26\%). We reason that this is likely due to merger-induced AGN fueling in our sample. 
	\item 35 broad H$\alpha$ galaxies feature BL emission offset from the galaxy center, outside the galaxy's 2.$^{\prime\prime}$5 nuclear region; 77 for the broad H$\beta$ galaxies. 3 galaxies with BL emission offset from the galaxy center feature both broad H$\alpha$ and H$\beta$ emission, leading to 109 unique galaxies with offset BL emission in our sample, where the emission is solely from the companion. For J005620.79+135143.6, we identify broad H$\beta$ emission in two companion galaxies.
\end{enumerate}
BL detection in large spectroscopic surveys offers a powerful tool for identifying AGNs, and for comprehending the AGN physics closest to the accretion disk, which is essential for unraveling the nature of AGN-galaxy co-evolution. 

Future work to enhance these efforts could include a thorough investigation of the kinematic properties of the BLR. Exploring the kinematics of the BL profiles in more depth (e.g., measuring the asymmetries of the line profiles and analyzing double-peaked BLs) can shed insight into how various BL velocity components manifest. These include, but are not limited to, a blend of Doppler motions, inflows/ outflows (possibly induced by mergers), shocks, rotation, or even a recoiling black hole. To further evaluate the kinematics of the BLs and to determine their relationship with feedback, constructing additional velocity maps (similar to the flux maps) for each BL galaxy would help to reveal the kinematic gradients of the gas (0.$^{\prime\prime}$5 x 0.$^{\prime\prime}$5 resolution; provided by MaNGA's DAP). This could help differentiate shocks from photoionized regions (e.g., shock velocities $\ge 500$ km s$^{-1}$). Moreover, outflowing winds can be identified by inspecting asymmetric and broadened Gaussian profiles.
\label{sec:summarybl}
\acknowledgments
\section*{Acknowledgments}
\textbf{J.N. and J.M.C. acknowledge support from NSF AST1714503 and NSF AST1847938.}

Funding for the Sloan Digital Sky Survey IV has been provided by the Alfred P. Sloan Foundation, the U.S. Department of Energy Office of Science, and the Participating Institutions. SDSS-IV acknowledges
support and resources from the Center for High-Performance Computing at
the University of Utah. The SDSS website is www.sdss.org.

SDSS-IV is managed by the Astrophysical Research Consortium for the 
Participating Institutions of the SDSS Collaboration including the 
Brazilian Participation Group, the Carnegie Institution for Science, 
Carnegie Mellon University, the Chilean Participation Group, the French Participation Group, the Harvard-Smithsonian Center for Astrophysics, 
Instituto de Astrof\'isica de Canarias, The Johns Hopkins University, Kavli Institute for the Physics and Mathematics of the Universe (IPMU) / 
University of Tokyo, the Korean Participation Group, Lawrence Berkeley National Laboratory, 
Leibniz Institut f\"ur Astrophysik Potsdam (AIP),  
Max-Planck-Institut f\"ur Astronomie (MPIA Heidelberg), 
Max-Planck-Institut f\"ur Astrophysik (MPA Garching), 
Max-Planck-Institut f\"ur Extraterrestrische Physik (MPE), 
National Astronomical Observatories of China, New Mexico State University, 
New York University, University of Notre Dame, 
Observat\'ario Nacional / MCTI, The Ohio State University, 
Pennsylvania State University, Shanghai Astronomical Observatory, 
United Kingdom Participation Group,
Universidad Nacional Aut\'onoma de M\'exico, University of Arizona, 
University of Colorado Boulder, University of Oxford, University of Portsmouth, 
University of Utah, University of Virginia, University of Washington, University of Wisconsin, 
Vanderbilt University, and Yale University.
Collaboration Overview
Start Guide
Affiliate Institutions
Key People in SDSS
Collaboration Council
Committee on Inclusiveness
Architects
Survey Science Teams and Working Groups
Code of Conduct
Publication Policy
How to Cite SDSS
External Collaborator Policy

This publication makes use of data products from the Wide-field Infrared Survey Explorer, which is a joint project of the University of California, Los Angeles, and the Jet Propulsion Laboratory/California Institute of Technology, funded by the National Aeronautics and Space Administration. This research has made use of data supplied by the UK Swift Science Data Centre at the University of Leicester.
 \software{Astropy \citep{2013A&A...558A..33A, 2018AJ....156..123A}}, IRAF \citep{1986SPIE..627..733T}.

\bibliographystyle{aasjournal}	
\bibliography{refs}	


\appendix
\label{appendixa}
\section{BL Catalog}	
\label{appendixfff}
\begin{table}[ht]
	\caption{Broad H$\alpha$ Galaxies} 
	\renewcommand{\thetable}{\arabic{table}}
	\centering
	\begin{tabular}{cccccc}
		\hline
		\hline
  		SDSS Name &  Plate-IFU & \textit{z}  & FWHM & L$_{\rm{BL}}$  & SMBH Mass \\ {} & {}& {} & {(km s$^{-1}$)} & {(10$^{38}$ erg s$^{-1}$)} & {(10$^{4}$ M$_{\odot}$)} \\
		\hline
J000131.42+142426.8 & 8075-6101 & 0.042 & 1105 $\pm$ 60 & 2.14 $\pm$ 0.1 & 1.29 $\pm$ 0.33 \\
J000625.57+152543.4 & 12700-6102 & 0.044 & 1666 $\pm$ 114 & 1.93 $\pm$ 0.12 & 2.84 $\pm$ 0.68 \\
J002820.10-001304.1 & 8656-12702 & 0.061 & 2066 $\pm$ 137 & 2.66 $\pm$ 0.19 & 5.31 $\pm$ 1.14 \\

		\hline
		\multicolumn{6}{p{13cm}}
  {Note: Columns are (1) SDSS name, (2) MaNGA Plate-IFU, (3) host galaxy stellar continuum redshift, (4) median FWHM of the broad-line spaxels in the galaxy, (5) broad-line luminosity, and (6) SMBH mass for the broad-line galaxy. This table is available in its entirety in a machine-readable form in the online journal. A portion is shown here for guidance regarding its form and content.}
\end{tabular}
\label{tab:blha}
\end{table}

\begin{table}[ht]
	\caption{Broad H$\beta$ Galaxies} 
	\renewcommand{\thetable}{\arabic{table}}
	\centering
	\begin{tabular}{ccccccc}
		\hline
		\hline
  		SDSS Name &  Plate-IFU & \textit{z}  & FWHM & L$_{\rm{BL}}$  & SMBH Mass & R$_{\rm{BLR}}$\\ {} & {}& {} & {(km s$^{-1}$)} & {(10$^{38}$ erg s$^{-1}$)} & {(10$^{4}$ M$_{\odot}$)} & {($\rm{light-days}$)} \\
		\hline
J001548.79+153614.8 & 8088-3702 & 0.116 & 1140 $\pm$ 127 & 1.95 $\pm$ 0.19 & 3.91 $\pm$ 0.43 & 0.07 $\pm$ 0.04 \\
J001643.42+150909.1 & 8088-6102 & 0.117 & 1160 $\pm$ 114 & 4.86 $\pm$ 0.42 & 6.76 $\pm$ 0.63 & 0.13 $\pm$ 0.06 \\
J001938.78+144201.1 & 8088-6104 & 0.116 & 1207 $\pm$ 129 & 9.6 $\pm$ 0.9 & 10.71 $\pm$ 0.87 & 0.21 $\pm$ 0.09 \\
		\hline
		\multicolumn{7}{p{15cm}}
  {Note: Columns are (1) SDSS name, (2) MaNGA Plate-IFU, (3) host galaxy stellar continuum redshift, (4) median FWHM of the broad-line spaxels in the galaxy, (5) broad-line luminosity, (6) SMBH mass for the broad-line galaxy, and (7) the radius of the broad line region for the broad-line galaxy. This table is available in its entirety in a machine-readable form in the online journal. A portion is shown here for guidance regarding its form and content.}
\end{tabular}
\label{tab:blhb}
\end{table}

\clearpage

\end{document}